\documentclass[12pt,preprint]{aastex}
\def\msun{\rm M_{\sun}}

\def\lsun{\rm L_{\sun}}

\def\av{${\rm A_V}$}

\begin{document}
\shortauthors{Hernandez et al.}
\shorttitle{ {\em Spitzer} disk census in Ori OB1}

\title{{\em Spitzer} observations of the Orion OB1 association: disk census in the low mass stars}

\author{Jes\'{u}s Hern\'andez\altaffilmark{1,2}, Nuria Calvet\altaffilmark{1}, C. Brice\~{n}o\altaffilmark{2}, 
L. Hartmann\altaffilmark{1}, A. K. Vivas\altaffilmark{2}, J. Muzerolle\altaffilmark{3}, 
J. Downes\altaffilmark{2,4},  L. Allen \altaffilmark{5}, R. Gutermuth \altaffilmark{5}}

\altaffiltext{1}{Department of Astronomy, University of Michigan, 830 Dennison Building, 500 Church Street, Ann Arbor, MI 48109, US}
\altaffiltext{2}{Centro de Investigaciones de Astronom\'{\i}a, Apdo. Postal 264, M\'{e}rida 5101-A, Venezuela}

\altaffiltext{3}{Steward Observatory, University of Arizona, 933 North Cherry Avenue, Tucson, AZ 85721, US}

\altaffiltext{4}{Escuela de F\'{\i}sica, Universidad Central de Venezuela, Apdo. Postal 47586, Caracas 1041-A, Venezuela}

\altaffiltext{5}{Harvard-Smithsonian Center for Astrophysics, 60 Cambridge, MA 02138, US}

\email{hernandj@umich.edu}

\begin{abstract}
We present new {\em Spitzer Space Telescope} observations of two fields
in the Orion OB1 association. We report here IRAC/MIPS observations 
for 115 confirmed members and 41 photometric candidates of the $\sim$10 Myr 
25 Orionis aggregate in the OB1a subassociation, and 106 confirmed members 
and 65 photometric candidates of the 5 Myr region located in the OB1b subassociation.
The 25 Orionis aggregate shows a disk frequency of 6 \% while the field 
in the OB1b subassociation shows a disk frequency of 13 \%. 
Combining IRAC, MIPS and 2MASS photometry we place stars bearing disks 
in several classes: stars with optically thick disks (class II systems),
stars with an inner transitional disks (transitional disk candidates) 
and stars with ``evolved disks'';  the last exhibit smaller IRAC/MIPS 
excesses than class II systems. In all, we identify 1  transitional 
disk candidate in the 25 Orionis aggregate and 3 in the OB1b field;
this represents $\sim$10\% of the disk bearing stars, indicating that
the transitional disk phase can be relatively fast. 
 We find that the frequency of disks is a function 
of the stellar mass, suggesting a maximum around stars with  spectral type M0. 
Comparing the infrared excess in the IRAC bands
among several stellar groups we find that inner disk emission 
decays with stellar age, showing a correlation with the respective
disk frequencies. The disk emission at the IRAC and MIPS bands 
in several stellar groups indicates that disk dissipation 
takes place faster in the inner region of the disks. Comparison with models
of irradiated accretion disks, computed with several degrees of settling, suggests
that  the decrease in the overall accretion rate observed in young stellar groups
is not sufficient to explain the weak disk emission observed in the 
IRAC bands for disk bearing stars with ages 5 Myr or older; 
larger degrees of dust settling are necessary to explain these objects.

\end{abstract}


\section{Introduction}
\label{sec:int}

Observational and theoretical studies indicate that important processes in 
the evolution of  protoplanetary disks take place at ages between 1 and 10 Myr. 
About 90\% of the low mass stars ($\sim$K5 or later) have lost their 
primordial disks at 5-7 Myr \citep[e.g.,][]{haisch01, hartmann05b, hernandez07}.
Grains grow to sizes of $\sim$1000 km stirring up the leftover small objects 
in the disks and originating the first generation of reprocessed dust by 
collisional cascades \citep{kenyon05, hernandez06}. Giant planets  
are expected to form in this period \citep{pollack96, alibert04}.
However, additional studies of disk population in this crucial age range
are necessary to improve our knowledge and clarify many details 
about the evolution from primordial disks to planetary systems.

OB associations are excellent laboratories
for comparative studies of protoplanetary disk evolution, because they harbor 
young stellar populations (1-10 Myr) originating from the same giant molecular clouds, 
spanning a wide range of stellar masses, and in a variety of evolutionary 
stages and environments \citep{brown99,aurora06, briceno07a, preibisch07}.
In particular, the Orion OB1 association (Ori OB1), as other OB associations, 
shows a well defined age sequence suggesting a large-scale triggered 
star formation scenario \citep{briceno05, briceno07a, lee07}.
Ori OB1 contains very young subgroups 
(ages $\lesssim$ 1 Myr) still embedded in their natal gas 
\citep[e.g., Orion A and B clouds; ][]{megeath06}, subgroups in the process of dispersing 
their natal gas \citep[e.g., the $\sigma$ Orionis cluster, age $\sim$ 3 Myr; ][]{hernandez07} 
and more evolved populations, which have long since dissipated their progenitor molecular 
clouds \citep[e.g., the 25 Orionis aggregate, age$\sim$ 10 Myr; ][]{briceno07b}.

We are carrying out an optical photometric and spectroscopic survey of 
$\sim$ 128 $deg^2$ in Ori OB1 in order 
to identify the low and intermediate mass stellar populations, and study the properties 
linked to the first stages of star and disk evolution 
\citep{briceno01, briceno05, briceno07b, briceno07c, calvet05a, hernandez05, hernandez06,hernandez07}. 
In this work, we expand the results from the optical survey with the capabilities of the {\em Spitzer Space 
Telescope} at near and mid infrared wavelengths to identify and characterize 
protoplanetary disks around young stellar objects 
\citep[e.g.][]{allen04, megeath04, gutermuth04,muzerolle04, hartmann05,aurora06, hernandez07}.
In particular we study the near and mid infrared properties of stars 
in two IRAC/MIPS {\em Spitzer} fields encompassing 
an area of $\sim$2.5 deg$^2$. One field is located in the 
7-10 Myr 25 Orionis stellar aggregate \citep{briceno07b}, 
the most populous $\sim$ 10 Myr stellar group known within 500 pc; the other 
is located in the Ori OB1b sub-association, in which we have estimated an age 
of $\sim$ 5 Myr \citep{briceno05,briceno07b,hernandez06}. Additional results 
from the 3 Myr $\sigma$ Orionis cluster \citep{hernandez07}, 
also located in the Ori OB1b subassociation, allow us to cover most of 
the potentially crucial age range in protoplanetary disk evolution.
In this cluster, we found 336 photometric members using optical and near 
infrared color-magnitude diagrams, about a third of this sample exhibits 
excess in the IRAC and/or MIPS bands indicating that they have disks.

This paper is organized as follows. In \S \ref{sec:obs} we present the observational
data and a brief description about membership. We analyze the observations 
and describe the disk emission detection in \S \ref{sec:res}. The main results 
are  summarized in \S \ref{sec:conc}.

\section{Observations}
\label{sec:obs}

\subsection{Infrared photometry}
\label{sec2:ir}

We have obtained near-infrared (NIR) and mid-infrared photometry  of 
two regions in the Orion OB1 association using the four channels (3.6, 4.5, 5.8 \& 8.0 {\micron}) of the InfraRed Array Camera \citep[IRAC, ][]{fazio04}, 
and the 24 {\micron} band of the Multiband Imaging Spectrometer 
\citep[MIPS;][]{rieke04}, on board the {\em Spitzer Space Telescope}. 
The field located in the 25 Orionis aggregate (hereafter ``25 Orionis'') 
covers an area of $\sim$1.1 deg$^2$ centered at RA$\sim$ 5.42 hours and DEC$\sim$1.64 deg;
the other field (hereafter ``OB1b'') covers an area of $\sim$1.4 deg$^2$ 
on the Orion OB1b sub-association centered at RA$\sim$5.52 hours and DEC$\sim$-1.71 deg. 
Dust infrared emission maps \citep{schlegel98} reveal that at least 90\% 
of the regions covered by IRAC images in 25 Orionis and OB1b 
have visual extinctions smaller than {\av}$\sim$0.12 
and {\av}$\sim$0.6, respectively \citep[see][]{hernandez06}.
These values are mostly in agreement with the mean visual extinction 
calculated from individual stars in \citet{briceno05}.

The IRAC observations were done using a standard raster map with 290" offsets,
to provide maximum areal coverage with just a slight overlap between frames,
to aid in mosaicking the data.  Each position is composed of 3 dithers,
with a single-frame integration of 12 seconds. The IRAC observations were processed 
using the {\sl IRACproc} \citep{schuster06} package to create the final mosaics 
with a scale of 0.86 \arcsec/pixel \citep[see][]{hernandez06}. Point source detections 
were carried out individually on each IRAC channel using PhotVis tool (an IDL 
GUI-based photometry visualization tool developed by R. Gutermuth). More than 
20,000 sources in each field were detected in at least one {\em Spitzer} band.
We extracted the photometry of these objects using the {\it apphot} package in IRAF,
with an aperture radius of 3\arcsec.7 and a background annulus from 3.7 to 8\arcsec.6.
We adopted zero-point magnitudes for the standard aperture radius (12\arcsec) and
background annulus (12-22\arcsec.4) of 19.665, 18.928, 16.847 and 17.391 in the
[3.6], [4.5], [5.8] and [8.0] channels, respectively. Aperture corrections were 
made using the values described in IRAC Data Handbook \citep{reach06}.

MIPS observations were obtained using the medium scan mode with full-array
cross-scan overlap, resulting in a total effective exposure time per pointing
of 40 seconds.  The images were processed using
the MIPS instrument team Data Analysis Tool (DAT), which calibrates
the data and applies a distortion correction to each individual exposure
before combining it into a final mosaic \citep{gordon05}. 
We obtained point source photometry at 24 {\micron} with IRAF/{it daophot} 
point spread function fitting, using an aperture size of about 5.7" and
an aperture correction factor of 1.73 derived from the STinyTim PSF model.
The absolute flux calibration uncertainty is less than 5\%.
Our final flux measurements are complete down to about 1 mJy in both maps 
(the limit flux is about 0.5 mJy).

\begin{figure}
\plotone{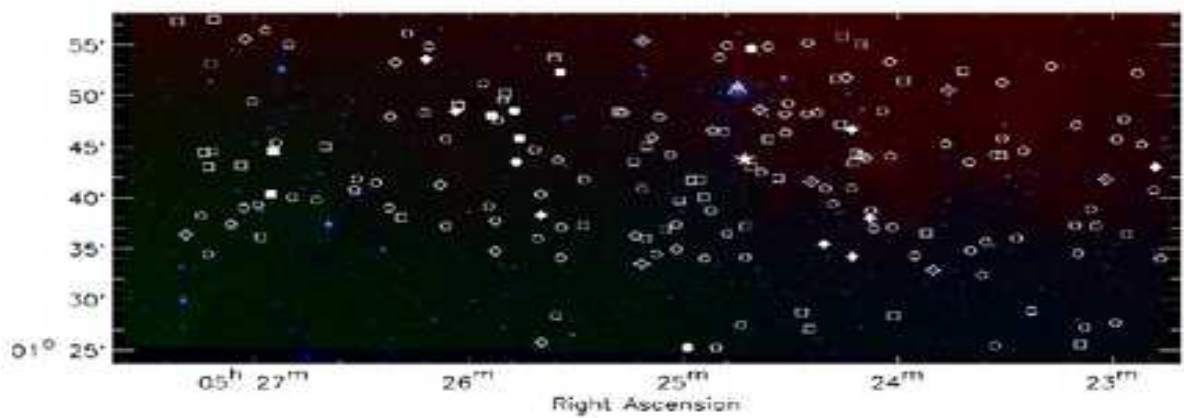}
\caption{IRAC image of the 25 Orionis aggregate. This is a 3-color composite of IRAC 
images (3.6 (blue), 4.5 (green), and 8.0 (red) \micron) illustrating the space 
distribution of members in this aggregate. 
Circles represent stars confirmed as low mass members using spectroscopic 
data \citep{briceno05,briceno07b,briceno07c}, squares represent 
photometric low mass candidates of 25 Orionis. Solid symbols indicate stars 
bearing  disks (see \S \ref{s:disksel}). Intermediate mass members 
selected in \citet{hernandez06} are represented 
as diamonds, open symbols represent stars without disks and solid symbols represent 
debris disk candidates.  The star indicates the Herbig Ae star V346 Ori \citep{hernandez06} 
and the triangle indicates the Be star 25 Ori.[See electronic edition of {\apj} for a color version of this figure.] }
\label{f:map1a}
\end{figure}

\begin{figure}
\plotone{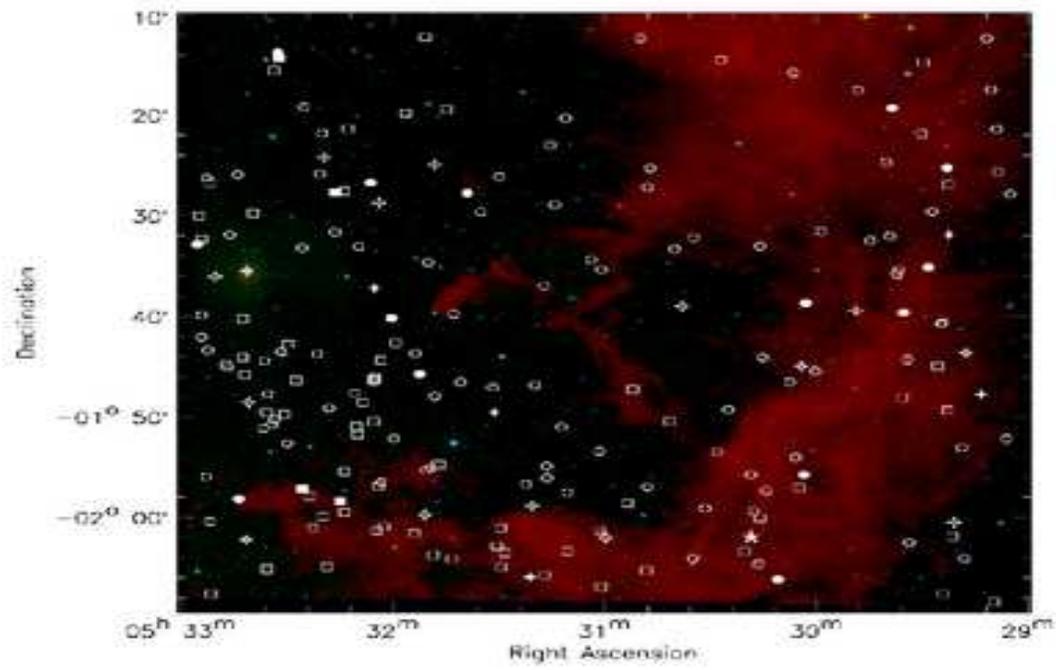}
\caption{IRAC image of the OB1b region. This is a 3-color composite of IRAC images,
3.6 (blue), 4.5 (green), and 8.0 (red) \micron. Symbols are as in Figure \ref{f:map1a}. 
The star indicates the Herbig Ae star HD290543 \citep{hernandez06}[See electronic edition of {\apj} for a color version of this figure.]}
\label{f:map1b}
\end{figure}

Figures \ref{f:map1a} and  \ref{f:map1b} show color images combining three channels 
of IRAC ([3.6],[4.5] and [8.0]) for 25 Orionis and for Ori OB1b, respectively.
We display the low mass spectroscopic members from \citet{briceno05,briceno07b,briceno07c} 
and the low mass photometric candidates selected in \S \ref{s:mem}; 
the stars bearing disks studied in \S \ref{s:disksel};  
and  the intermediate mass members including the debris disk candidates and 
the Herbig Ae stars studied in \citet{hernandez06}.

\subsection{Optical photometry}
\label{sec2:opt}
Optical (V and I Cousin) magnitudes were obtained from the CIDA Variability Survey
which is being carried out using the QUEST I camera \citep{baltay02} 
installed on the Jurgen Stock Telescope (a  celar aperture 1-m Schmidt telescope) 
at the Venezuela National Astronomical Observatory. The camera, an array of 4x4 CCDs, 
is designed to work in driftscan mode, which is a very efficient way to 
survey large areas of the sky. Each scan was reduced and calibrated with the 
standard QUEST software and the method described in  \citet{vivas04}, 
in which variable stars can be identified \citep[see ][]{briceno05}.

\subsection{Low mass members and photometric candidates}
\label{s:mem}

We follow the procedures described in \citet{hernandez07} to reject 
non stellar objects and contaminating sources using IRAC 
color-color and IRAC color-magnitude diagrams. In brief,  we select 
stars with [3.6]$<$14.5, below this limit, the contamination  from 
extragalactic sources is expected to be more than 50\% \citep{fazio04}. 
The [4.5]$-$[5.8] versus [5.8]$-$[8.0]  and [3.6]$-$[5.8] versus 
[4.5]$-$[8.0] color-color diagrams were used 
to eliminate most of the galaxies with polycyclic aromatic 
hydrocarbon (PAH) emission and objects with strong 
8 \micron contamination (Gutermuth et al. 2007, in preparation).

Optical and 2MASS counterpart \citep{cutri03} for the IRAC sources 
were found using a 2'' matching radius. A preliminary list 
of 623 objects in 25 Orionis and 918 objects in OB1b were created 
using optical-2MASS color magnitude diagrams (V versus V-Ic, 
V versus V-J and J versus J-K)  to select  those objects above the
zero age main sequence \citep[ZAMS; ][]{sf00} at the distance of each 
stellar group \citep[330 pc and 440 pc for 25 Orionis and OB1b, respectively; ][]{briceno05,briceno07b,hernandez05}.
We rejected by visual inspection non-members sources, like diffuse objects, and 
objects with  an apparent problem in the photometry, like close binaries, faint companion 
binaries, and stars on the image border. 
 
Low mass members were confirmed by \citet{briceno05, briceno07b, briceno07c} using
optical spectra obtained at the Fred Lawrence Whipple Observatory 
with the 1.5 m telescope equipped with the FAST spectrograph \citep{fabricant98},
and with the 6.5 m MMT telescope equipped with the  Hectospec and Hectochelle multifiber 
spectrographs \citep{fabricant05}.  Additional low resolution spectra were
obtained at the Kitt Peak National Observatory using the Hydra multiobject 
spectrograph on the 3.5 m WIYN telescope. Low mass members of 25 Orionis 
and Ori OB1b can be identified by the presence of Li I $\lambda$6707
in absorption  and the equivalent width of the H$\alpha$ line in emission \citep[e.g.][]{briceno98}. 
The strength of the H$\alpha$ line is used to separate accreting stars, 
the Classical T Tauri stars (CTTS), from non-accreting stars, the Weak T Tauri stars (WTTS).
To distinguish between the two type of objects 
we followed the prescription adopted by \cite{white03}. 
Spectral types were derived using the SPTCLASS code 
\footnote{http://www.astro.lsa.umich.edu/\~hernandj/SPTclass/sptclass/html}, 
which includes several molecular features 
(like TiO and VO bands), characteristic of low mass stars. Some stars 
in our sample have additional spectroscopic confirmation using radial velocity 
distributions. We refer the reader to \citet{briceno05,briceno07b,briceno07c}
for more details about spectroscopic membership confirmation of stars belonging to
the Ori OB1 association. Tables \ref{t:mem1a} and \ref{t:mem1b} show the IRAC and MIPS 
photometry for 115 confirmed members of 25 Orionis and 106 confirmed members of
Ori OB1b. Column 1 shows the internal running identification number 
in each sample, columns 2  shows the 2MASS designation \citep{cutri03}; 
columns 3 and 4 are the stellar coordinates; columns 5, 6, 7 and 8 give IRAC 
magnitudes in the bands [3.6], [4.5], [5.8] and [8.0], respectively; column 9 
gives MIPS (24 \micron) magnitudes; column 10  gives the reference
for the optical spectroscopic and photometric data 
; the last column gives the disk classification based on the IRAC 
and MIPS analysis (\S \ref{s:disksel}).

\begin{figure}
\epsscale{1.0}
\plotone{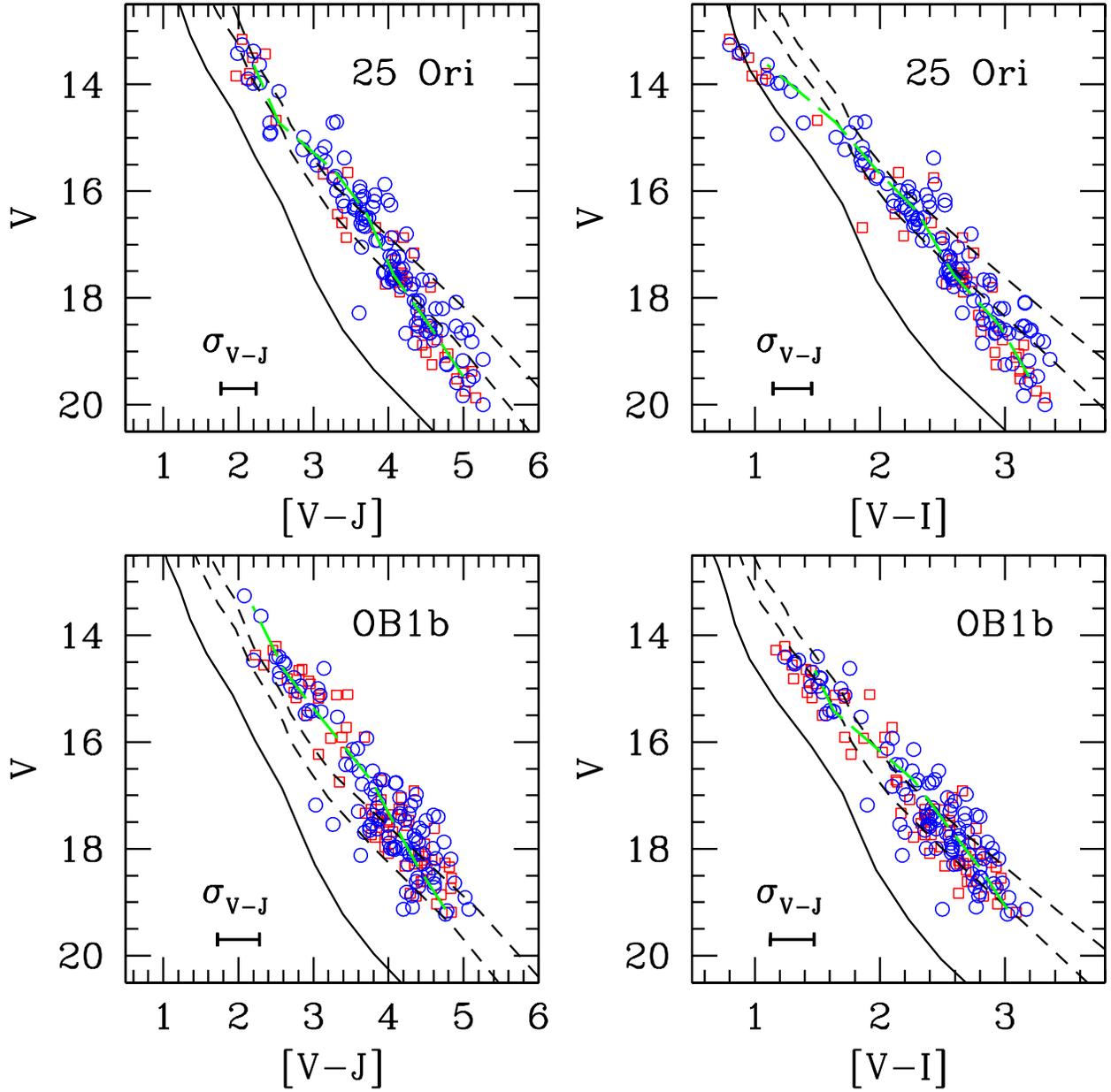}
\caption{Color magnitude diagrams illustrating the selection of low mass photometric 
candidates in 25 Orionis (upper panels) and in OB1b (lower panels). Photometric candidates 
(open squares)  have similar distribution than members (open circles) confirmed using 
spectroscopic methods \citep{briceno05,briceno07b, briceno07c}. 
By comparison, the ZAMS (solid lines) and the isochrones at 
10 Myr and 5 Myr \citep[ dashed lines;][]{sf00} are displayed 
at the distance of each stellar group; 330 pc for 25 Orionis and 440 pc for OB1b 
\citep{briceno05,briceno07b,hernandez05}. Long dashed lines represent the median 
colors for the spectroscopic members.[See electronic edition of {\apj} for a color version of this figure.]}
\label{f:cmd}
\end{figure}

Figure \ref{f:cmd} shows optical-2MASS color magnitude diagrams (CMDs), V vs V-I 
and V versus V-J, illustrating  the selection of the photometric candidates 
of 25 Orionis  (upper panel) and Ori OB1b ( lower panels). Solid lines represent
the ZAMS \citep{sf00} for each stellar group. Dashed lines represent the 
10 Myr  and 5 Myr isochrones \citep{sf00}. Since, it is well-known that theoretical, 
non-birthline isochrones do not so well for intermediate mass stars 
\citep{hartmann03} and the opacity tables at low temperatures  are incomplete 
\citep{lyra06, baraffe98}, these theoretical isochrones are plotted 
as reference  and do not affect the selection of the photometric candidates.
We assumed a distance of 330 pc for 25 Orionis and 440 pc for OB1b \citep{briceno05,briceno07b, hernandez05}.
The distribution of confirmed members (open circles) can be roughly traced as a 
straight line in the CMD diagrams, with a larger spread in Ori OB1b due 
to the larger visual extinction in this region in comparison with 25 Orionis. 
We calculated the median colors (V-I and V-J) for confirmed members (open circles) 
within 1.0 mag bins in the visual band. We used the differences between the observed colors
and the expected colors (the median; represented by long dashed lines) 
to calculate the standard deviation ($\sigma$) of the member samples.
We selected as photometric candidates (open squares) stars located 
in the region described by the 2.5$\sigma$ limits in the CMDs \citep[see ][]{hernandez07}.  
Tables \ref{t:phot1a} and \ref{t:phot1b} show the IRAC, MIPS  and optical photometry 
for the 41 photometric candidates found in 25 Orionis and for the 65 candidates found 
in OB1b, respectively. 
The information shown in these tables in columns from 1 to 9 is
the same as in Table \ref{t:mem1a}. Columns 10 and 11 show the V magnitude 
and the color V-I from the CIDA survey; the last column gives the 
disk classification based on  the IRAC and MIPS analysis (\S \ref{s:disksel}).

\section{Results}
\label{sec:res}

\subsection{Disk diagnostics}
\label{s:disksel}

\begin{figure}
\epsscale{1.0}
\plotone{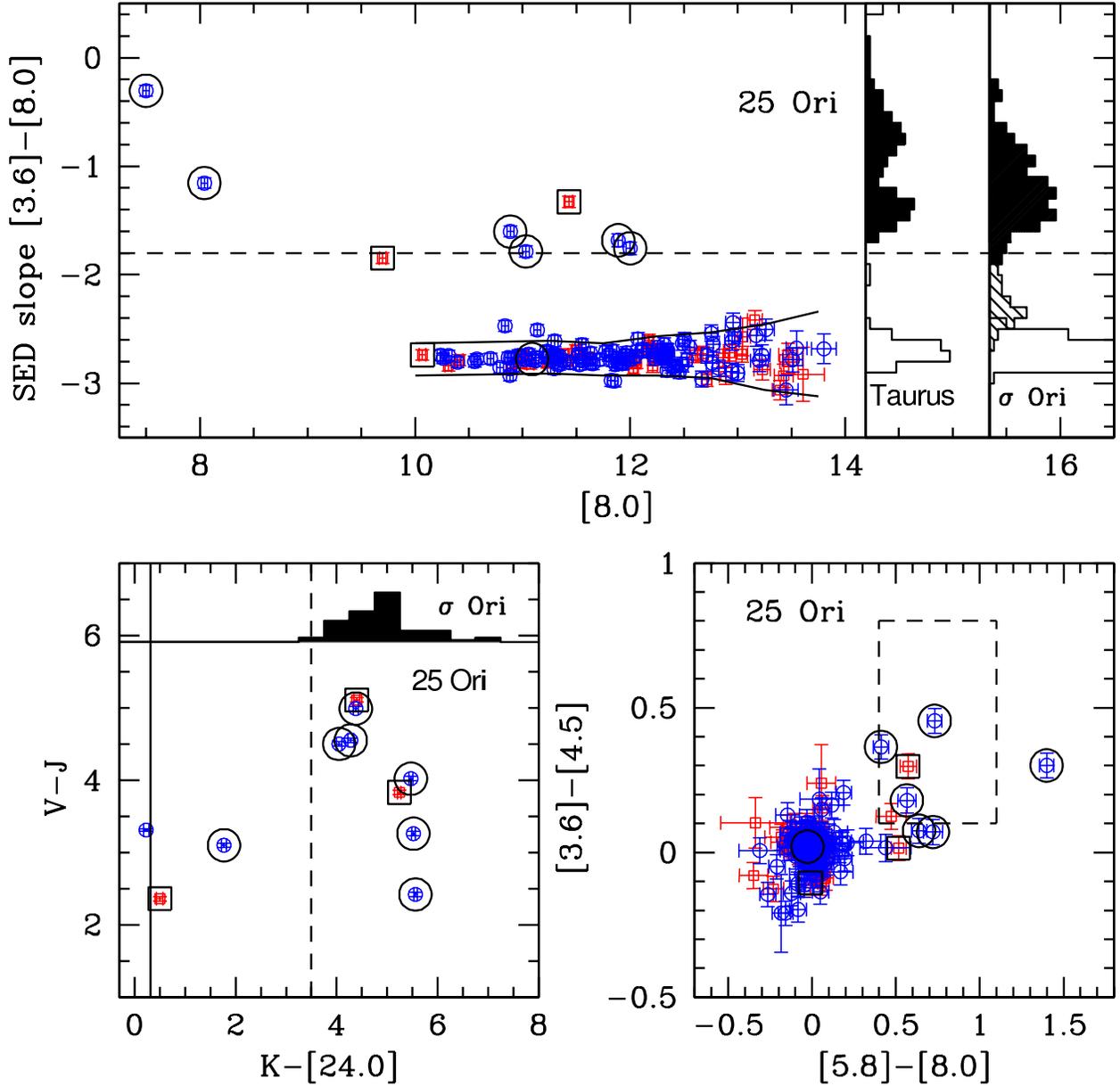}
\caption{Diagrams illustrating the detection of sources with infrared excess 
above the photospheric levels (solid lines) in  25 Orionis.
The top panel shows the IRAC SED slope diagram indicating stars 
with excess at 8 \micron.
The bottom-left panel shows a 2MASS-MIPS color-color diagram indicating stars 
with excess emission at the MIPS 24 {\micron} band. The bottom-right panel shows an 
IRAC color-color diagram indicating objects with infrared excess at [4.5] 
and [8.0]. Members and photometric candidates are displayed 
as circles and squares, respectively. Large circles and large squares 
represent stars with infrared excess at 24 \micron. The solid histograms 
represent the distribution of IRAC SED slopes for stars with optically 
thick disks in Taurus \citep{hartmann05} and in the $\sigma$ Orionis cluster \citep{hernandez07}; 
dashed lines represent the class II region limit based on these histograms. 
The open histograms represent the distribution of stars with no IRAC excesses. 
The dashed histogram is the distribution of stars with evolved disks in the 
$\sigma$ Orionis cluster \citep{hernandez07}[See electronic edition of {\apj} for a color version of this figure.]}
\label{f:disk1a}
\end{figure}

\begin{figure}
\epsscale{1.0}
\plotone{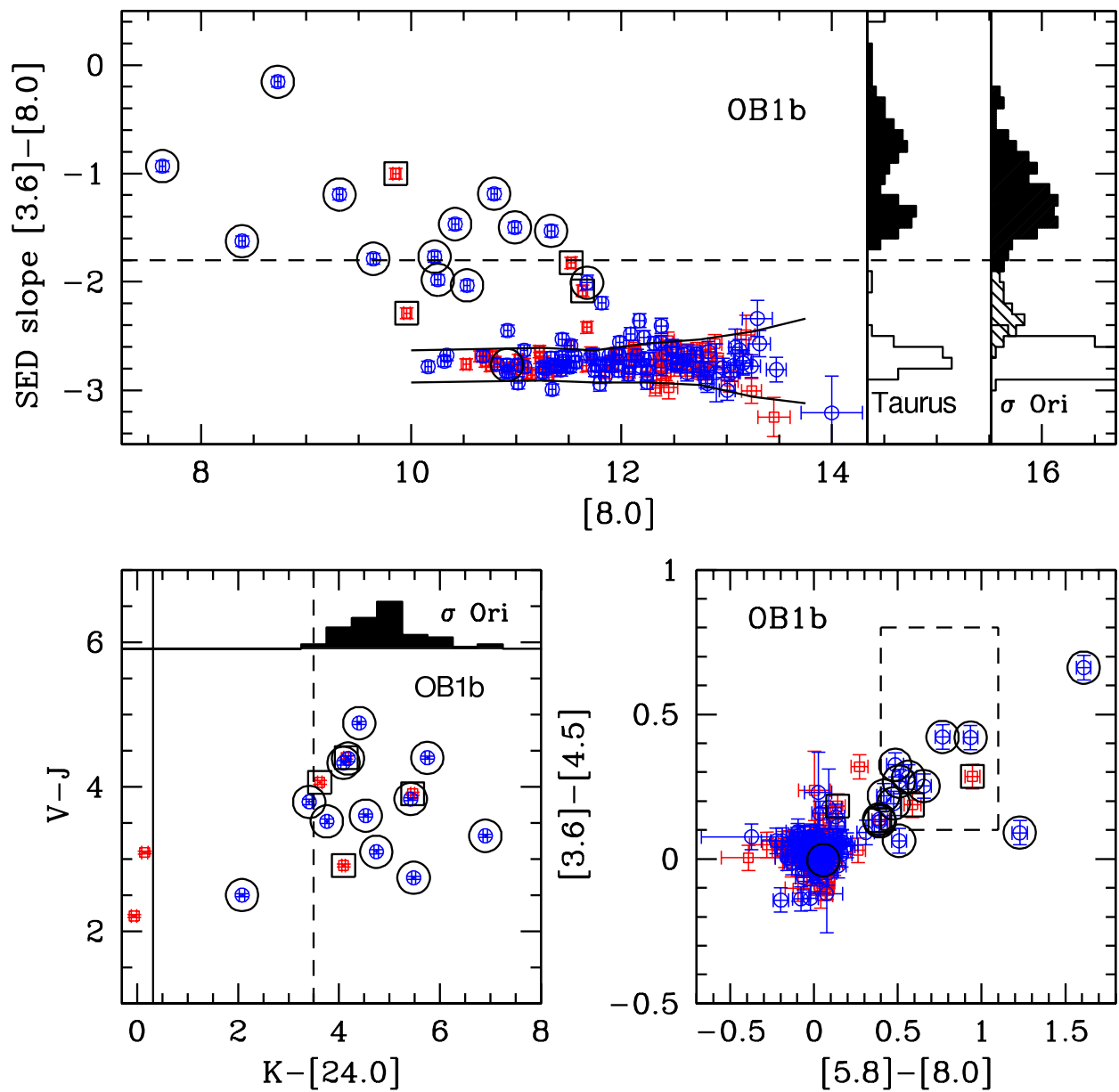}
\caption{Diagrams illustrating the detection of sources with infrared excess 
above the photospheric levels (solid lines) in the OB1b field. 
Symbols are similar to Figure \ref{f:disk1a}.[See electronic edition of {\apj} for a color version of this figure.]}
\label{f:disk1b}
\end{figure}

Figures \ref{f:disk1a} and \ref{f:disk1b} show three diagrams used to identify 
and roughly characterize the stars bearing disks in 25 Orionis and OB1b, 
respectively.   
The top panels  show the SED slope, determined from the [3.6]$-$[8.0] color (IRAC SED slope), 
versus the [8.0] magnitude for members (open circles) and photometric candidates (open squares). 
The photospheric level is described by the upper solid line, which is calculated using the photometric 
errors propagated from the [3.6]$-$[8.0] color \citep{hernandez07}. Stars with excess emission at 
8 {\micron} can be identified in this diagram.  For comparison, IRAC SED slope histograms for Taurus \citep{hartmann05}
and for the $\sigma$ Orionis cluster  \citep{hernandez07} are displayed. In general, disk bearing stars in Taurus (solid histogram) 
are located  in a well defined region (which we call the class II region) with an IRAC SED slope $>$ -1.8 \citep[see ][]{lada06}; 
this limit ( dashed lines)  is used to identify objects with optically thick disks 
in which the inner disk emission has not been affected  significantly by evolutionary processes. 
In contrast, 15\% of the disk bearing stars in the $\sigma$ Orionis cluster
exhibit smaller IRAC excesses (dashed histogram) suggesting a 
reduction in disk photosphere height, possibly due to dust growth
and/or settling \citep{hernandez07}.
The bottom left panel shows the K$-$[24] versus V-J color-color diagram, 
in which we identify  members (big open circles) and photometric candidates 
(big open squares) with 24 {\micron} infrared emission above the photospheric level 
(solid lines) indicating that disks are present around these objects \citep[e.g.; ][]{gorlova06, hernandez06,hernandez07}.
In this panel, we display the K$-$[24] color distribution for stars bearing disks in the $\sigma$ 
Orionis cluster with an IRAC SED slope $>$ -1.8 (which represents a disk
population similar to those found in Taurus) and we use this histogram to identify
stars with K$-$[24] color characteristic of stars with optically thick disks ( K$-$[24]$>$3.5, class II region).
In the upper panel and in the bottom left panel, we define the ``evolved disk region''
between the class II region and the photospheric region. 
The bottom right panel shows the IRAC color-color diagram, in which 
we identify stars with excess emission in the IRAC bands 
\citep[e.g. ][]{allen04,megeath04,hartmann05, aurora06, hernandez07}.
The dashed box displays the colors predicted for CTTS 
of different accretion rates by the models of \citet{dalessio05b}. 
In general, the IRAC colors observed for disk bearing stars
in Taurus are located in this region \citep{hartmann05, aurora06}.

 In the top panel of Figure \ref{f:disk1a}, we identify six members and two photometric
candidates located on the IRAC class II region; most of them are  located near 
the class II limit possibly indicating that these objects have begun the process 
of clearing the inner primordial disk.  Two members  have very small IRAC excesses 
just above the photospheric region. These objects, also located between the photospheric and 
the CTTS regions in the IRAC color-color diagram, 
have no MIPS detections and therefore it is not clear if the small excess observed at 
8 {\micron} originates from PAH background contamination, by an unresolved companion, 
or by disks present around these stars (flagged as ``disk[8]?'' in Table \ref{t:mem1a}).
Of particular interest are the member 1a\_1121 and the photometric  candidate 
1a\_1626 which are located between the photospheric and 
the class II region in the V-J versus K$-$[24] diagram indicating that the outer disks 
around these objects are in a more evolved stage.  Moreover these stars are 
also located on the photospheric region in the IRAC color-color diagram and in the IRAC SED 
slope diagram indicating that the inner disk has  already dissipated and no disk 
emission can be detected at wavelength $\lesssim$8 \micron. The star 1a\_1626 also 
has a very small excess at 24 {\micron} ( $\sim$ 2 $\sigma$ above the photospheric level)
indicating that the presence of a disk around this object is not yet conclusive.
Overall, in 25 Orionis we identify 7 stars with disks in the member sample 
(disk frequency 6.1$\pm$2.3\%), and 3 in the photometric candidate sample 
(disk frequency 7.3$\pm$4.2\%).

Similarly, in the top panel of  Figure \ref{f:disk1b} we identify 13 members 
and 4 photometric candidates in Ori OB1b that show IRAC and MIPS excesses; 
five of these objects are located between the class II region and the photospheric region.
Eight members and one photometric  candidate with no MIPS detection are located 
in the evolved disk region (flagged as ``disk[8]?'' in Table \ref{t:mem1b} and \ref{t:phot1b}). 
The existence of disks around these objects needs additional confirmation 
since they could be below the MIPS detection limit, or could be contaminated by PAH 
background emission (in Figure \ref{f:map1b} it can be clearly seen that 
the sky background emission at 8 {\micron} is very patchy, and significant at some locations).
In general, the range of infrared excesses at 24 {\micron} in OB1b is 
similar to that of the optically thick disks in the $\sigma$ Orionis 
cluster. Only one disk bearing star (6\%), the star 1b\_337, has 24 {\micron} excess 
below the class II limit while 6 stars (39\%) have 8 {\micron} excess below this limit.  
This suggests a more rapid decrease in dust emission in the inner disk, in agreement
with results from \citet{aurora06} in the Cepheus OB2 association.
The member 1b\_337, located in the evolved disk region in the V-J versus K$-$[24]
diagram, does not have excess in the IRAC bands. 
 Overall, in Ori OB1b we identify 14 stars with disks in 
the member sample (disk frequency 13.1$\pm$3.5\%) and 
the 4 disk systems in the photometric candidate sample (disk frequency 6.2$\pm$3.1\%).

\begin{figure}
\epsscale{1.0}
\plotone{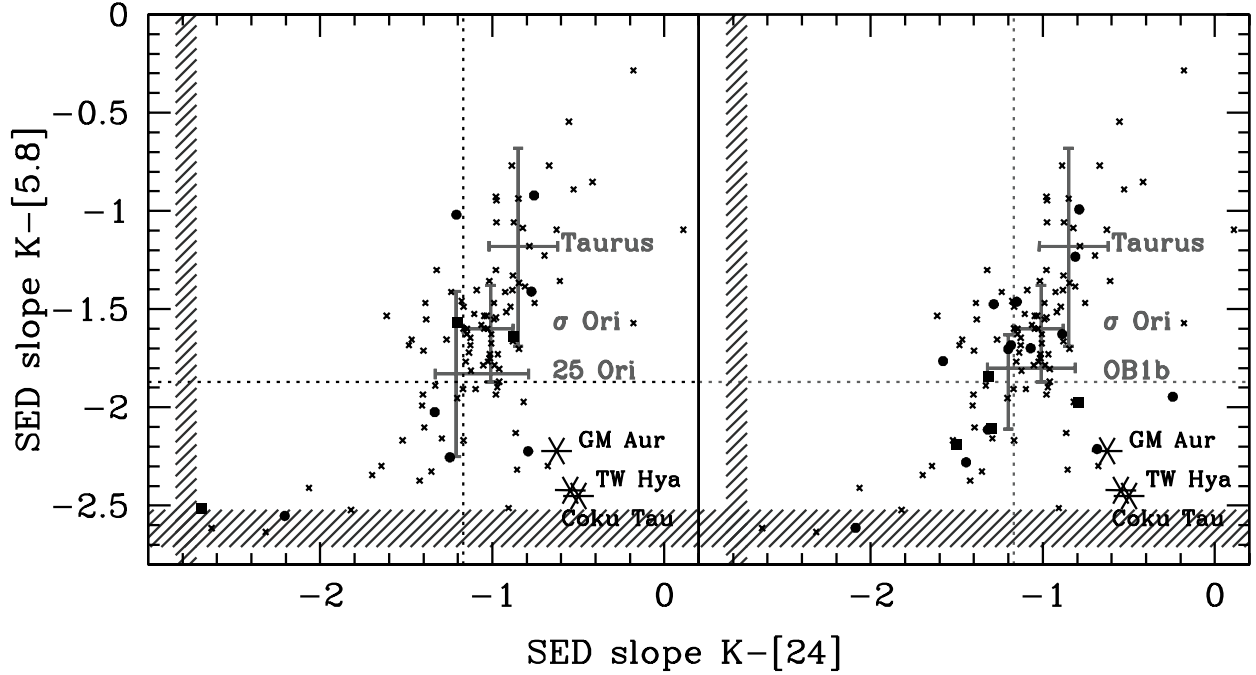}
\caption{SED slope K$-$[5.8] versus SED slope K$-$[24] for members (circles) 
and photometric candidates (squares) bearing disks in 
25 Orionis (left panel) and in  OB1b (right panel). 
Error bars represent the quartiles of the stellar groups, 
Taurus (1 Myr), the $\sigma$ Orionis cluster (3 Myr), OB1b  (5 Myr), and 
25 Orionis (10 Myr). Stars represent the transitional disk objects, Coku Tau\/4, TW Hya, and GM Aur 
\citep{dalessio05a,uchida04, calvet02,calvet05b} . 
Photospheric regions derived for K5-M5 stars using STAR-PET {\em Spitzer} tool are 
represented as dashed areas. Disk bearing stars in the
$\sigma$ Orionis cluster \citep{hernandez07} are represented with the symbol X. 
The horizontal dotted line represents the lower quartile of the stars bearing disks 
in the $\sigma$ Orionis cluster, $\sim$96 \% of stars with disks in Taurus are 
located above this line; we define the area above this line as 
the class II region (CII in Table 1-4).
The vertical dotted line represents the lower quartile of the stars bearing disks 
in the $\sigma$ Orionis cluster. We use this limit to separate transitional disk
candidate (TD in Table 1-4) from stars with evolved disks (EV in Table 1-4).}
\label{f:slope}
\end{figure}

Figure \ref{f:slope} displays the distribution of the disk bearing stars in a SED slope space
diagram for 25 Orionis (left panel) and OB1b (right panel). The vertical axis is 
the SED slope calculated from the K$-$[5.8] color and the horizontal axis  is the SED slope 
calculated from the K$-$[24] color.  The dashed areas define 
the photospheric level calculated with the STAR-PET {\em Spitzer} tool for stars K5 or later.
By comparison the low mass stars bearing disks of the 3 Myr  
$\sigma$ Orionis cluster \citep{hernandez07} are also plotted. Error bars represent 
the quartiles of disk bearing stars in Taurus, calculated from the K$-$[5.8] color 
in \cite{hartmann05} and from the K$-$[24] color estimated from the
median SED slope in \cite{furlan06}, in the $\sigma$ Orionis cluster from \cite{hernandez07},
in 25 Orionis and in Ori OB1b from this work. 
An overall decrease in the infrared excess 
is observed from the 1 Myr old  stars in Taurus to the 5 and 10 Myr old stars studied in 
this work; the disk population of the 3 Myr $\sigma$ Orionis cluster represents
an intermediate stage in this evolution.   Moreover, it is apparent  that the decrease
is larger at 5.8 {\micron} than at 24 {\micron}, indicating that evolution processes 
occur faster in the inner region of the disk. 

In order to characterize the stars bearing disks in 25 Orionis and OB1b, we 
identify several regions in Figure \ref{f:slope} defined by the dotted lines 
(``class II region'', ``evolved disk region'' and ``transitional disk region'').
The horizontal dotted line represents the lower quartile of the 
$\sigma$ Orionis cluster. Above this line $\sim$96\% of the stars bearing 
optically thick disks in Taurus are also located, indicating a 
limit where the inner disk emission has not been affected significantly 
by evolutionary process. We can identify the  class II objects as 
stars located above this line. In general, the class II objects 
identified using Figure \ref{f:slope} are located in the class II 
region in Figures \ref{f:disk1a} and \ref{f:disk1b}.
Disk bearing stars below the dotted lines have decreased the disk 
infrared emission at 5.8 {\micron} due to a decrease in the irradiation 
surface of the inner disks, and so they are in an stage 
where processes for inner disk dissipation have begun. 
 The vertical dotted line represents the lower quartile of the 
stars bearing disks in the $\sigma$ Orionis cluster 
(the lower quartile of Taurus is rightward from this line). 
Using this limit, the stars located below the dotted lines
could be sub-grouped based on their disk emission at 24 \micron: 
``evolved disks objects'' ( SED slope K$-$[24] $\la$  -1.2), 
in which we see an overall decrease in the disk emission in 
the IRAC and MIPS bands, 
indicating similar evolution in the inner and outer disk \citep{lada06,hernandez07}; 
and ``transitional disk candidates'' ( SED slope K$-$[24] $\ga$  -1.2), which have 
an inner optically thin disk region, combined with an outer, optically 
thick disk \citep[e.g.;][]{calvet05b}. As reference, we plotted 
3 transitional disk stars, Coku Tau/4 \citep{dalessio05a}, TW Hya \citep{calvet02, uchida04}, 
GM Aur \citep{calvet05b}, which occupy the region defined for the 
``transitional disk candidates''.  In brief, we identify 5 stars with optically 
thick disks (class II objects), one transitional disk candidate and 4 evolved disk
objects in  25 Orionis. We also identify 10 class II objects, 
3 transitional disk candidates and 5 evolved disk objects in OB1b.
In spite of the evolved disks and transitional disks objects being a subsequent stage from 
class II objects, it is not clear if transitional disk objects are a pre-stage 
of evolved disks or each stage represents an independent stage from class II objects.

\begin{figure}
\epsscale{1.0}
\plotone{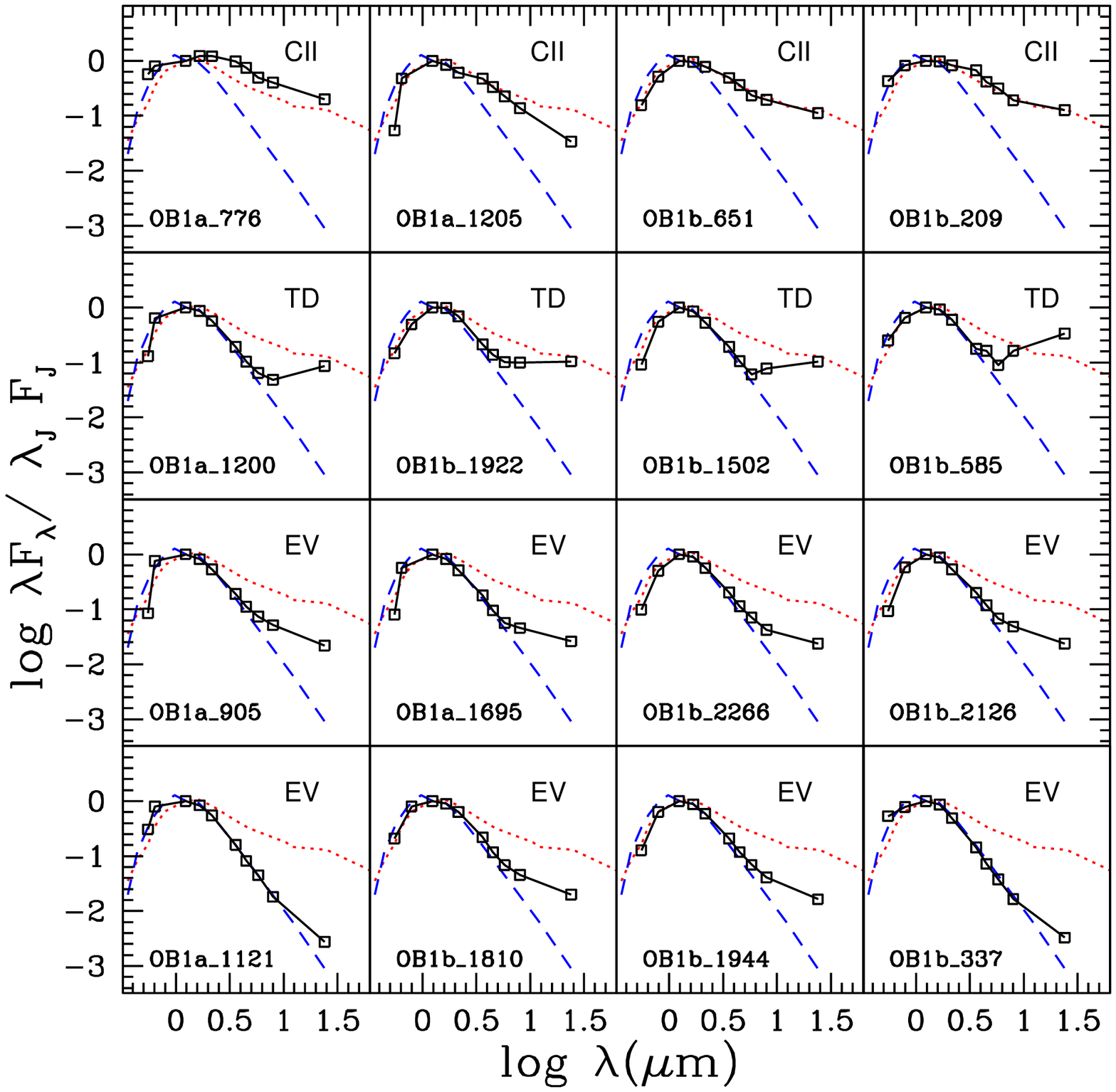}
\caption{Spectral energy distributions for selected disk bearing stars, illustrating the diversity of disks found 
at 5 and 10 Myr. Dotted line represents the median SEDs of disk bearing stars in Taurus \citep{hartmann05}.
Dashed line represents the median photosphere of stars in the spectral type range K5-M5 \citep{kh95}. We display 
examples of the disk types found in Figure \ref{f:slope}: stars with optically thick disks (CII), transitional disk 
candidates (TD) and evolved disks objects (EV).[See electronic edition of {\apj} for a color version of this figure.]
}
\label{f:seds}
\end{figure}

Figure \ref{f:seds} shows SEDs for selected stars in our samples illustrating the 
disk classification based on Figure \ref{f:slope}. The first row of SEDs shows 
stars with optically thick disks (CII) located above the dotted line in Figure \ref{f:slope}. 
The second row shows transitional disk candidates (TD) located right and below the dotted 
lines in Figure \ref{f:slope}. Finally, last two rows of panels show stars 
with evolved disks located left and below the dotted lines in Figure \ref{f:slope}.

\subsection {Models}
\label{s:model}

We have calculated SED slopes for models of irradiated accretion disks including dust 
settling from \citet{dalessio06}. In these models the disk is assumed to be steadily 
accreting at rates of {\.{M}} = 1e-9, 1e-8, and 1e-7 M$\sun$/yr, 
onto a star with mass of 0.6 $\msun$
and luminosity of 1.2 $\lsun$, which corresponds to a K7 star with age of 1 Myr \citep{sf00}. 
Dust settling was included using two populations of grains (big and small grains)
having different spatial distributions, with the larger grains concentrated
toward the midplane. The small grains located in the upper layers
have different depletions given by the $\epsilon $ parameter (with values= 1, 0.1, 0.01, 0.001), 
which is the ratio of the dust to gas mass ratio of small grains relative to the
the standard dust to gas mass ratio \citep[$\zeta_{small}$/$\zeta_{std}$; ][]{dalessio06}.
The inner wall of the disk, located at the dust destruction radius, was settled 
self-consistently with the same degree of depletion used in the outer disk.

\begin{figure}
\epsscale{1.0}
\plotone{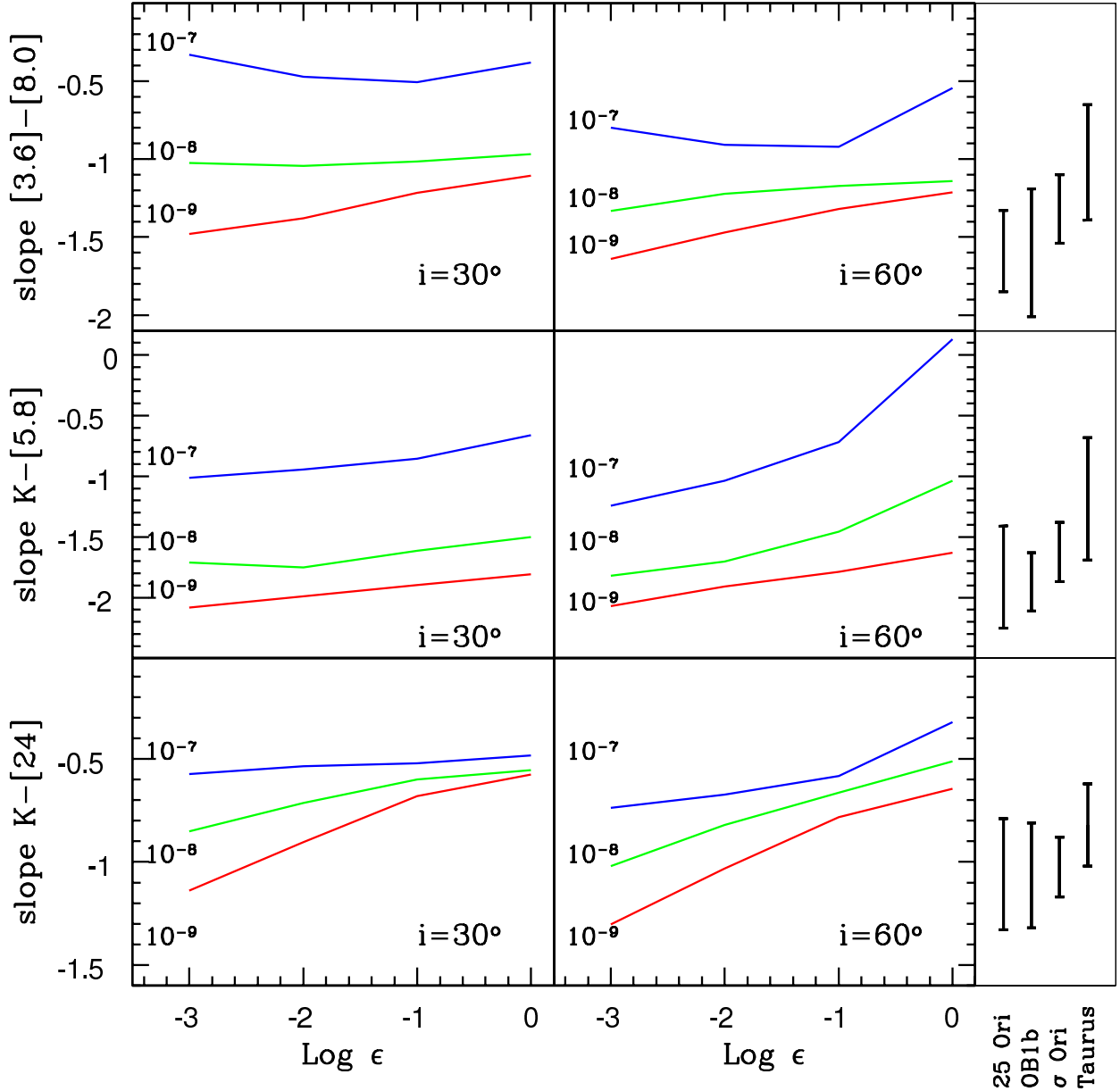}
\caption{Theoretical SED slopes as function of the degree of settling.
Each panel shows three curves corresponding to different 
accretion rates ({\.{M}}=$10^{-7}$,{\.{M}}=$10^{-8}$ and  {\.{M}}=$10^{-9})$.
The parameter $\epsilon$ represents the degree of settling in each model.
The central object is a typical T Tauri star with M=0.6$\msun$, 
Age=1 Myr, corresponding to a spectral type K7 \citep{sf00}.
Left and middle panels represent disks oriented with an angle of 
30$\deg$ and  60 $\deg$, respectively. In the right panels, we 
display the quartiles of the disk population in Taurus \citep[1-2 Myr;][]{hartmann05,furlan06},
the $\sigma$ Orionis cluster \citep[$\sim$3 Myr][]{hernandez07}, and the stellar groups
studied in this contribution, OB1b ($\sim$ 5 Myr) and 25 Orionis ($\sim$ 10 Myr). [See electronic edition of {\apj} 
for a color version of this figure.]}
\label{f:model}
\end{figure}

Figure \ref{f:model} shows the theoretical SED slopes derived from the colors, 
[3.6]$-$[8.0], K$-$[5.8] and  K$-$[24] versus the degree of settling represented by 
 $\epsilon$. SED slopes were calculated convoluting the theoretical SED with 
the transmission curves of the respective filters. We plot two inclination angles
along the line of sight , 30 deg (left panels) and 60 deg (middle panels); this 
range in angles represents 40\% of probability of observation. 
Accretion rates are indicated for the different curves plotted in each panel. 
The slope has a strong dependency on  {\.{M}}, showing flatter slopes for the 
fastest accretors; the smallest variation in disk emission with {\.{M}}
is observed for the slope K$-$[24] of disks without settling ($\epsilon$=1). 
In general, models with {\.{M}} = 10$^{-9}$ $\msun$/yr show a stronger dependence 
with dust settling than models for large accretion rates.

By comparison, we plotted in the right panels of Figure \ref{f:model} the 
quartiles observed for disk bearing stars in Taurus, 
in the $\sigma$ Orionis cluster, in Ori OB1b and in 25 Orionis.  
The range of disk emission observed in Taurus ( 1-2 Myr) can 
be explained by the models, indicating optically thick disks systems with several 
degrees of settling \citep{furlan06} and accretion rates \citep{hartmann98, calvet05a}. 
Most of the stars with disks in the $\sigma$ Orionis cluster ($\sim$80\%)
can be explained by the theoretical SED slopes but with small accretion rates 
or/and higher degree of dust settling than in Taurus. 
Approximately half of the disks observed in 25 Orionis (Figure \ref{f:disk1a}) 
and OB1b (Figure \ref{f:disk1b}) require models with lower accretion 
rates ({\.{M}} $<$ 10$^{-9}$ $\msun$/yr) or/and  large degree of 
settling ($\epsilon$$<$0.001) to explain the weak disk emissions
observed at [3.6]$-$[8.0] and K$-$[5.8]. However, $\sim$75\% 
of disk bearing stars in 25 Orionis  and OB1b have disk emissions 
at K$-$[24] in agreement with the SED slopes predicted by the models,
supporting the scenario where the inner disk evolves faster than the 
outer disk.

\subsection {Disk frequencies}
\label{s:diskfrec}

In Figures \ref{f:disk1a}, \ref{f:disk1b} and  \ref{f:slope}
we identify 7 members bearing disks in 25 Orionis and 14 in Ori OB1b, 
indicating disks frequencies in the member samples of 6.1$\pm$2.3\% 
and 13.1$\pm$3.5\%, respectively. These frequencies include objects 
with 24 {\micron} excess as disk bearing stars. In \S \ref{s:disksel}, 
we also identify 2 members of 25 Orionis and 8 members of Ori OB1b 
that exhibit IRAC 
excesses but have no MIPS detections; if we add these stars as members with
disks, the disk frequencies increase to 7.8$\pm$2.6\% in 25 Orionis 
and to 20.6$\pm$4.4\% in Ori OB1b. These later values are in better agreement 
with the disk frequencies calculated for the low mass stars in the 
Ori OB1a (7 $\pm$ 3 \%)  and Ori OB1b (17 $\pm$ 4 \%) sub associations, using 
the excess emission (2$\sigma$ above the expected photospheric level) from the 
2MASS color H-K \citep{hernandez05}.
The lower disk frequencies derived using the 
24 {\micron} excess could indicate a possible observational bias 
produced by the flux limit of MIPS observations ($\sim$ 0.5 mJy). 
Assuming the distances and ages \citep{sf00} 
plotted in Figure \ref{f:cmd}, with a visual extinction of \av=0.12 mag for 
25 Orionis and \av=0.6 mag for OB1b, we cannot expect to 
detect disks around a 0.6$\msun$ star in the 25 Orionis group if 
E$_{[24]} \lesssim$2.5, and if E$_{[24]} \lesssim$4.5 for objects in
Ori OB1b; where the excess ratio, E$_{[24]}$, 
is the ratio of the observed flux to the expected  photospheric flux at 24 \micron.

\begin{figure}
\epsscale{1.0}
\plotone{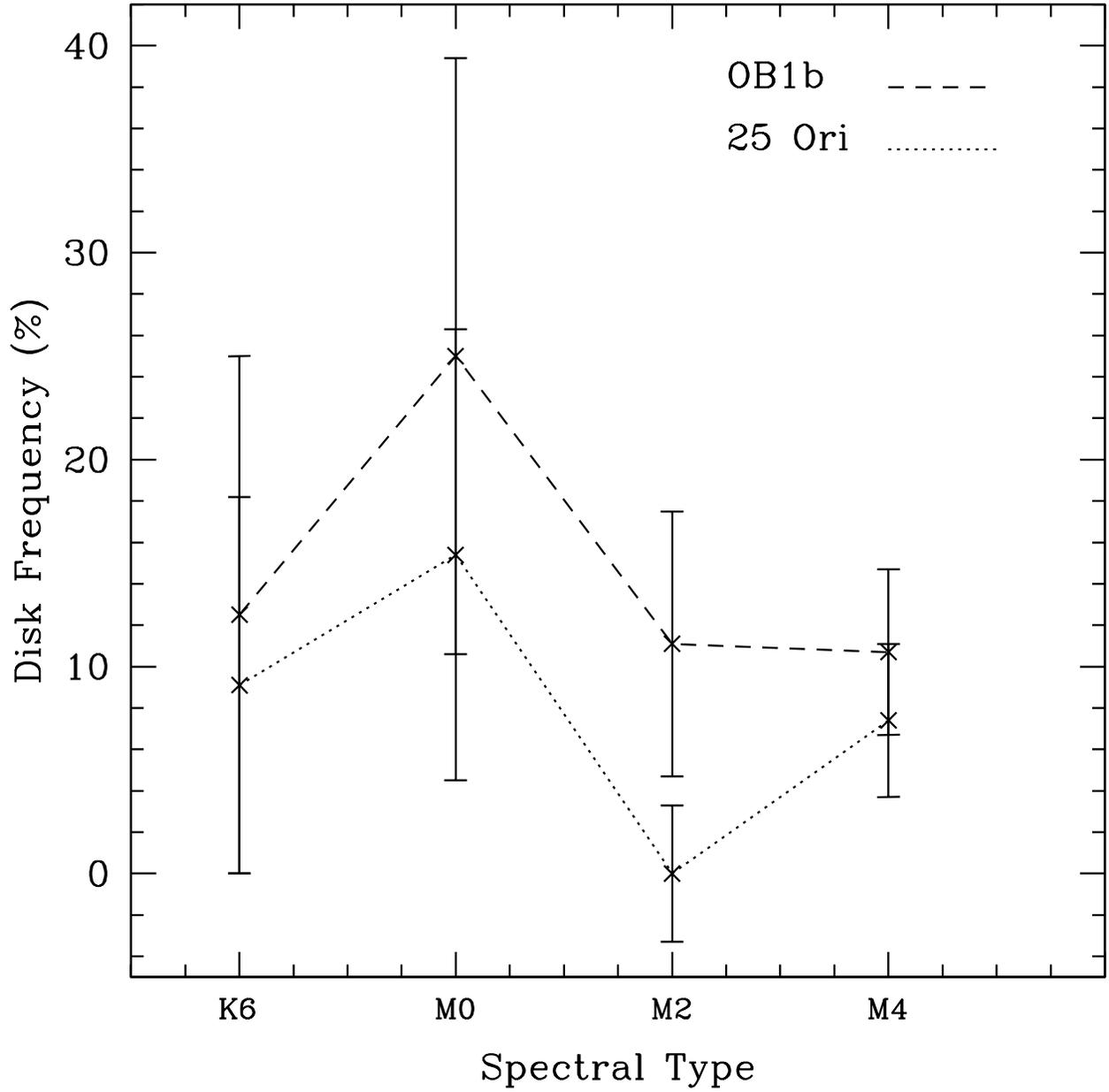}
\caption{Disk frequency as function of spectral type for 
the 25 Orionis aggregate (dotted line) and the OB1b field 
(dashed line).  The error bars represent the Poissonian 
statistical uncertainties}
\label{f:diskfrac}
\end{figure}

Using the stars identified as members with infrared 
excess in Figure  \ref{f:slope}, we plotted in Figure \ref{f:diskfrac}
the fraction of stars bearing disks versus  spectral type  
for 25 Orionis (dotted line) and OB1b (dashed line). Error bars represent 
the statistical $\sqrt{N}$ errors in our derived frequencies.
Previous studies have indicated that the frequency of disks is strongly dependent 
on the stellar mass, showing larger frequencies in the TTS mass range (spectral types K and M)
than in stars with higher masses \citep{lada95, aurora05, hartmann05b, 
hernandez05, carpenter06, briceno07a}. 
 Figure \ref{f:diskfrac} also suggests that the disk frequency declines toward lower masses 
(spectral types later than M1) showing a maximum around K7-M1 stars in agreement with results 
for the 2-3 Myr cluster IC 348 by \cite{lada06}. However, given the degree of uncertainty 
in each individual point in Figure \ref{f:diskfrac}, caused by the small number of disk 
bearing stars in each spectral type bin, 
plus the observational bias introduced by the limiting magnitude in 
the MIPS photometry, this result is not conclusive and additional 
data is necessary to confirm this trend.

Using Figure \ref{f:slope}, the frequencies calculated for stars 
with evolved disks related to the total number of stars bearing disks in 
the $\sigma$ Orionis cluster (15.6$\pm$4.0 \%), in Ori OB1b 
(27.8$\pm$12.4 \%) and in 25 Orionis (40$\pm$20\%),
indicate a clear trend toward more evolved disks in older stellar groups.
The transitional disks candidates in these stellar groups are less 
frequent than evolved disks (8$\pm$3\% for the $\sigma$ Orionis cluster, 
17$\pm$10 \% for OB1b and 10$\pm$10\% for 25 Orionis), suggesting that 
the transitional disk phase is relatively fast or represents an 
independent and special stage in the disk clearing period of stars.

Disk frequencies of the sample members can be used to estimate the contamination level
of non-members in the photometric candidates samples assuming that the photometric 
candidates bearing disks are actual members of young stellar groups. 
Since the overall disk frequencies of low mass members (6.1 $\pm$ 2.1 \%) and  
low mass photometric candidates (7.3 $\pm$ 4.2 \%) in 25 Orionis are similar,
we can expect that photometric candidates sample have very low non-member contamination.
Since the disk frequency calculated in OB1b for the photometric candidates 
(6.2 $\pm$ 3.1\%) is lower than  the disk frequency of members ( 13.1 $\pm$ 3.5), we can expect 
that around 50\% of the photometric candidates are not members of OB1b.

\subsection {Disk evolution}
\label{s:diskevo}

\begin{figure}
\epsscale{1.0}
\plotone{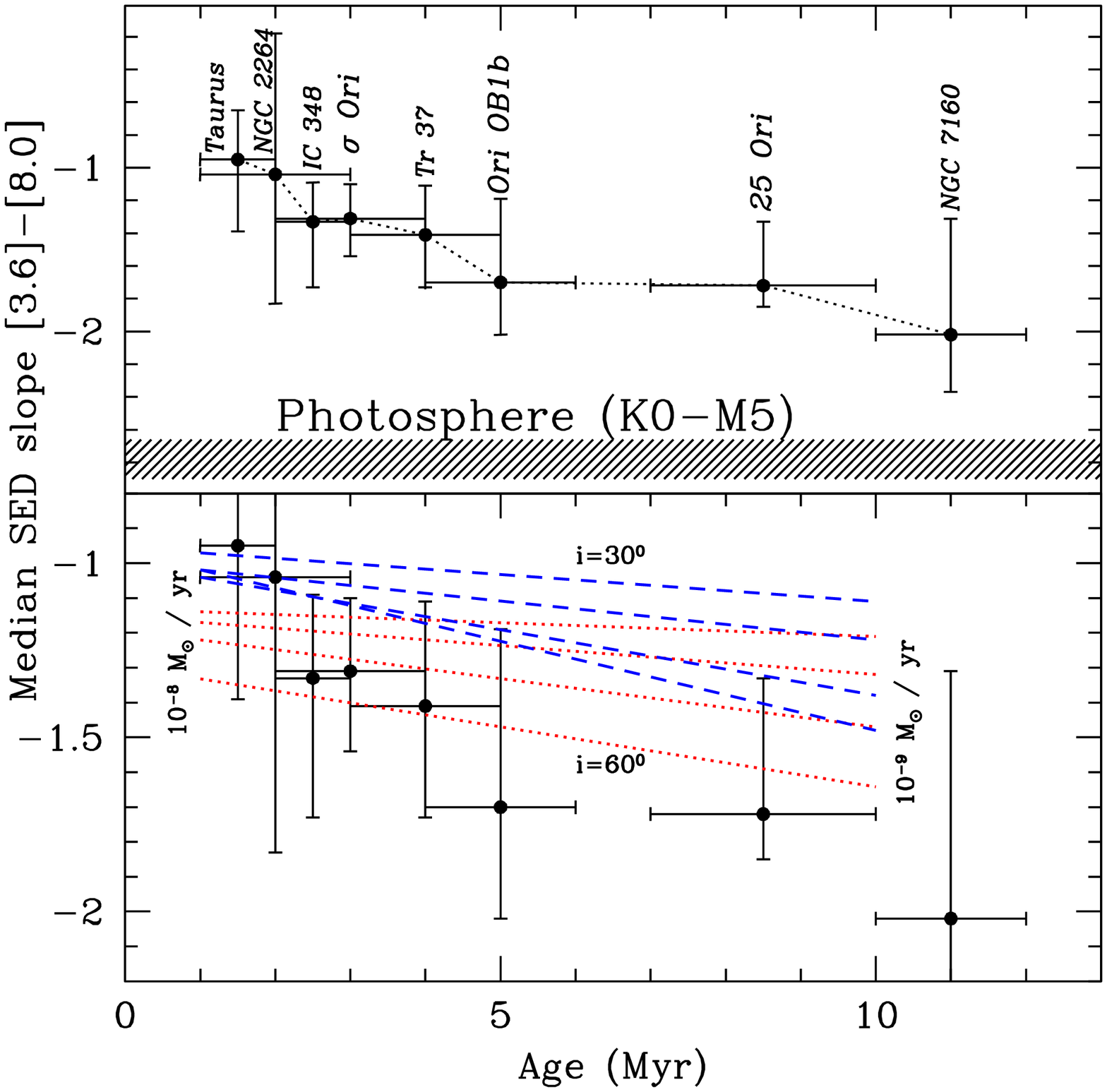}
\caption{Median SED slope in the IRAC spectral range versus stellar 
ages of different stellar groups: Taurus \citep{hartmann05},  
NGC 2264 \citep{young06}, IC 348 \citep{lada06},  $\sigma$ Orionis \citep{hernandez07}, 
Tr37 and NGC 7160 \citep{aurora06}, and 25 Orionis and OB1b 
\citep[][ and this work]{briceno07b}.
The dashed region indicates the photospheric levels estimated using STAR-PET {\em Spitzer} tool. 
It is apparent that the infrared disk emission decreases with the age of the stellar groups.
The lower panel shows a comparison with theoretical values described in \S \ref{s:model}. Two 
set of models are displayed showing different orientation of the disks, 30 deg (dashed lines)
and 60 deg (dotted lines). We assume canonical values for accretion rate at 1 Myr 
and 10 Myr (see text). Different lines describe different degree of settling ($\epsilon$=1,0.1,0.01 and 0.001) 
showing flatter slopes for models with less degree of settling.[See electronic edition of {\apj} for a color version of this figure.] }
\label{fig:diskevol}
\end{figure}

The frequency of stars bearing disks in different stellar groups is 
a function of age, indicating a time scale for disk 
dissipation in low mass stars of 5-7 Myr \citep[e.g.,][]{haisch01,briceno07a,hernandez05,hernandez07}.
The results presented in \S \ref{s:diskfrec} are in agreement with this trend.
In addition to this decline in disk frequency,
the amount of infrared disk emission also decreases with age. 
The top panel of Figure \ref{fig:diskevol} shows the median SED slope derived 
from the color [3.6]$-$[8.0]  for disk bearing stars in stellar groups 
ranging in age from $\sim$1 to $\sim$10 Myr: 
1-2 Myr \citep[Taurus;][]{hartmann05},
1-3 Myr  \citep[NGC 2264;][]{young06}, 2-3 Myr  \citep[IC348;][]{lada06}, 
3 Myr \citep[$\sigma$ Orionis;][]{hernandez07}, 4 Myr \citep[Tr 37;][]{aurora06}, 5 Myr 
\citep[OB1b;][ and this work]{briceno07b}, 7-10 Myr \citep[25 Orionis;][ and 
this work]{briceno07b} and 10-12 Myr \citep[NGC 7160;][]{aurora06}. We estimated 
the photospheric level using the STAR-PET {\em Spitzer} tool for the star templates 
with spectral types between K5 to M5.
Error bars represent the quartiles given the median value for each stellar group.
The median SED slope decreases with age, indicating  a reduction of disk emission
the inner regions of the disk. In the bottom panel, we display two sets of 
models described in \S \ref{s:model} with different disk orientations, 
i=30 $\deg$ (dashed lines) and i=60 $\deg$ (dotted lines). 
The accretion rate is a function of age \citep{hartmann98,muzerolle00,calvet05a}, so  
we assumed {\.{M}}=10$^{-8}\msun$ for 1 Myr old stellar groups, which is the mode  value
for accreting stars in Taurus \citep{hartmann98} and {\.{M}}=10$^{-9}\msun$ for 10 Myr
old stellar groups, which is the mode value for accreting stars in the 
OB1a sub-association \citep{calvet05a}.  
This plot suggests that the expected decrease in accretion rate 
is not sufficient to explain the decrease of emission in the 
inner part of the disk, and it is necessary to increase the 
degree of settling ($\epsilon$$<$0.001) in the inner disk 
to explain the observed slopes in stellar groups with ages 5 Myr or older.
 
\begin{figure}
\epsscale{1.0}
\plotone{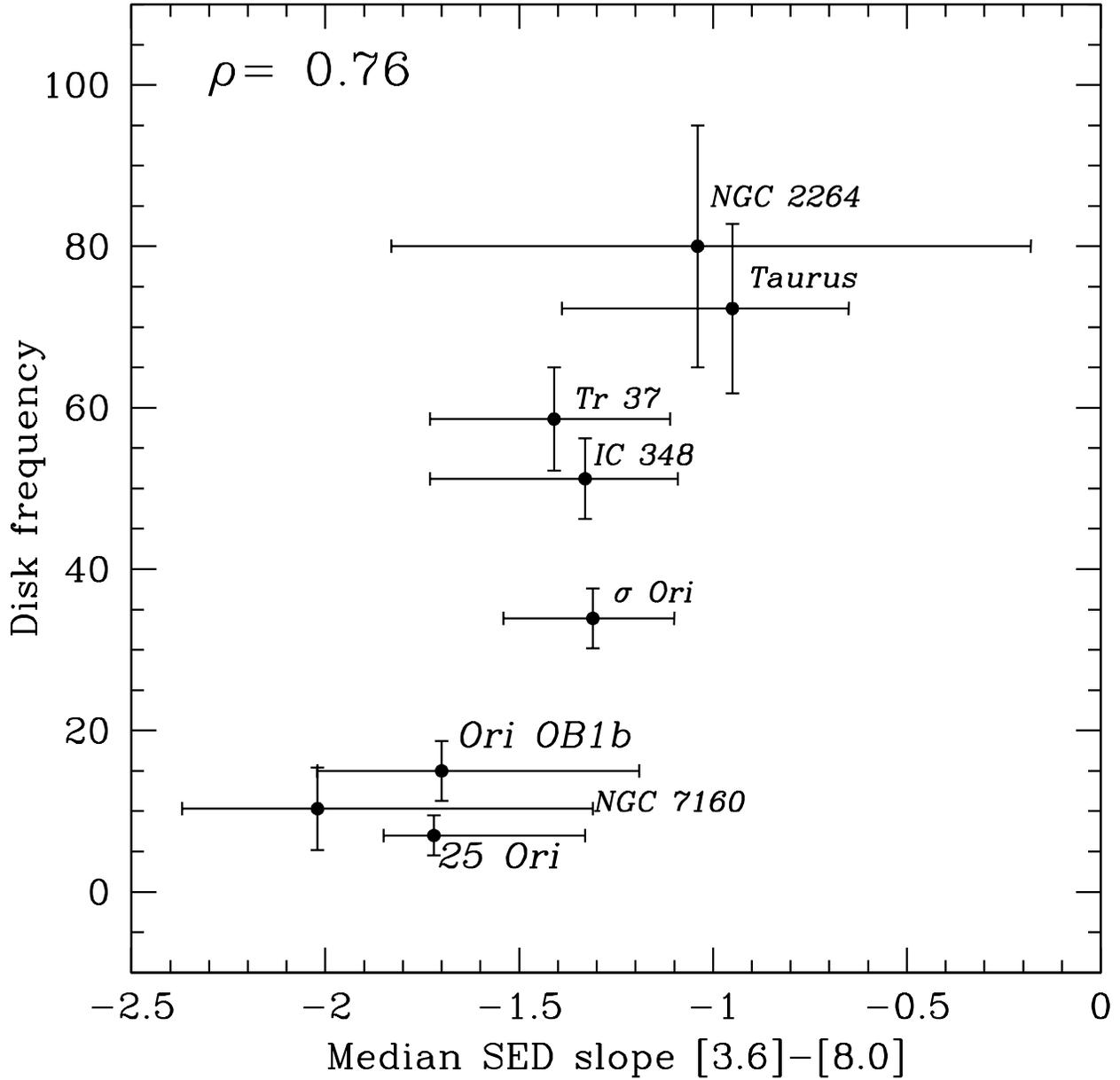}
\caption{IRAC disk Frequency versus the median SED slope [3.6]$-$[8.0] 
for different stellar groups (see Figure \ref{fig:diskevol}).
These measurements are correlated (correlation coefficient $\rho$=0.76) indicating more flaring disks 
in stellar groups with higher disk frequency.}
\label{fig:diskevol2}
\end{figure}

Since the disk frequency and the infrared disk emission decrease with age, 
a correlation between these values can be expected. Figure \ref{fig:diskevol2}
confirms this suggestion, showing that the disk frequency is correlated with 
the median SED slope in the IRAC bands (correlation coefficient $\rho$=0.76).
This plot clearly demonstrates that the disappearance of inner disks is related
to the 
decrease of optical depth in the inner disk due to the increase of dust
settling and the decrease of mass accretion rate \citep[see ][]{dalessio06}.

\section{Conclusions}
\label{sec:conc}

We have used the IRAC and MIPS data from {\em Spitzer} to study the disk frequencies 
and properties of disks around confirmed members of the Orion OB1 association, 
115 belonging to the $\sim 10$ Myr old 25 Orionis aggregate, 
and 106 in a region within the $\sim 5$ Myr old Ori OB1b sub association, 
near the Orion belt star $\epsilon$ Ori. 
Using optical-2MASS color magnitude diagrams 41 stars were selected 
as additional photometric candidates of  the 25 Orionis aggregate, 
and 65 as additional photometric candidates of the Ori OB1b field.
We use IRAC-MIPS diagrams to detect disk bearing stars, and to classify them
as either having no detectable disk emission (class III), as systems with 
optically-thick disks (class II), or as objects in an intermediate phase 
between class II and class III systems. These intermediate type objects 
were further grouped in two categories: 
``transitional disks candidates'', which have an inner, 
optically thin disk region combined with an outer, optically thick 
disk \citep[e.g.;][]{calvet02,calvet05b};  and ``evolved disk objects'', in 
which there is an overall decrease of emission both the inner 
and outer regions of the disk \citep{lada06,hernandez07}. It is not clear 
if ''evolved disks'' are a subsequent event after transitional disks phase, or
represent an independent evolutionary stage.

We found that the disk frequency in the 25 Orionis aggregate 
(6 \%) and  in the Ori OB1b field (13 \%)
is mass dependent, showing a maximum value for stars with spectral 
type M0, and suggesting a decrease in the disk frequency toward higher 
and lower masses. The trend toward higher masses has been observed in several 
stellar groups \citep[e.g.][]{hernandez05, aurora06, carpenter06}. The decrease in 
disk frequency toward lower masses is not conclusive, but is consistent
with the results for other regions like IC 348 \citep{lada06}.
We find that objects with evolved disks are more frequent in older 
stellar groups, while the transitional disk candidates represent a relative 
small fraction of the disk bearing stars in the various stellar groups,
suggesting that the transitional disk phase is relatively fast.

Comparing the disk emission in the IRAC and MIPS bands for Taurus, 
the $\sigma$ Orionis cluster, the Ori OB1b field and the 25 Orionis 
aggregate, we find that disk emission decreases faster 
in the innermost regions of the disk;
comparison with disk models from \citet{dalessio06} support 
this scenario. 
Finally, comparing the disk emission in the IRAC spectral range 
of several stellar groups ranging in age from $\sim$1 to 
$\sim$12 Myr , we find that inner  disk emission 
decreases systematically with age, showing a correlation between 
disk frequencies and inner disk emission. 
Comparison with models using a typical 
accretion rate for 1 and 10 Myr suggests that
viscous evolution alone
is not sufficient to explain the decrease in the inner 
disk emission, and that large degrees of dust settling ($\epsilon$ $<$ 0.001) 
are necessary to explain the observed SEDs at ages 5 Myr or older.

\acknowledgements

We thank Massimo Marengo for his advise during the reduction and mosaicking
of the IRAC data. This publication makes use of data products from
the CIDA Equatorial Variability Survey, obtained with
the J. Stock telescope at the Venezuela National Astronomical
Observatory, which is operated by CIDA for the Ministerio del
Poder Popular para la Ciencia y Tecnolog{\'\i}a of Venezuela, 
and from the Two Micron All Sky Survey, 
which is a joint project of the University
of Massachusetts and the Infrared Processing and Analysis Center/California
Institute of Technology.
This work is based on observations made with the {\em Spitzer Space Telescope} 
(GO-1 3437), which is operated by the Jet Propulsion Laboratory, California 
Institute of Technology under a contract with NASA. Support for this work 
was provided by  NASA grants NAG5-9670 and NAG10545.





\begin{deluxetable}{llllllllllllll}
\rotate
\tabletypesize{\tiny}
\tablewidth{0pt}
\tablecaption{Members of the 25 Orionis aggregate \label{t:mem1a}}
\tablehead{
\colhead{OB1a} & \colhead {2MASS} & \colhead{RA(2000)} & \colhead{DEC(2000)} & \colhead{[3.6]} & \colhead{[4.5]} & \colhead{[5.8]} & \colhead{[8.0]} & \colhead{[24.0]} & \colhead{Ref} & \colhead{Disk}\\
\colhead{ID} & \colhead{ID} & \colhead{deg} & \colhead{deg} & \colhead{mag} & \colhead{mag} & \colhead{mag} & \colhead{mag} & \colhead{mag} & \colhead{opt} & \colhead{types} \\
}
\startdata
9  &  05224654+0134010  &  80.69393  &  1.56697  &  12.419 $\pm$ 0.030  &  12.629 $\pm$ 0.131  &  12.275 $\pm$ 0.036  &  12.458 $\pm$ 0.042  &  \nodata $\pm$ \nodata  &        3       &  CIII \\
25  &  05224842+0140438  &  80.70176  &  1.67885  &  12.817 $\pm$ 0.030  &  12.777 $\pm$ 0.031  &  12.753 $\pm$ 0.041  &  12.824 $\pm$ 0.057  &  \nodata $\pm$ \nodata  &       3       &  CIII \\
47  &  05225186+0145132  &  80.71609  &  1.75367  &  13.269 $\pm$ 0.031  &  13.262 $\pm$ 0.032  &  13.139 $\pm$ 0.050  &  13.450 $\pm$ 0.112  &  \nodata $\pm$ \nodata  &       3       &  CIII \\
53  &  05225304+0152151  &  80.72102  &  1.87088  &  12.416 $\pm$ 0.030  &  12.348 $\pm$ 0.031  &  12.240 $\pm$ 0.036  &  12.258 $\pm$ 0.043  &  \nodata $\pm$ \nodata  &       3       &  CIII \\
905  &  05245885+0125183  &  81.24523  &  1.42177  &  12.904 $\pm$ 0.031  &  12.724 $\pm$ 0.031  &  12.456 $\pm$ 0.037  &  11.889 $\pm$ 0.040  &  9.20 $\pm$ 0.04  &    3       &  EV \\
930  &  05250192+0134563  &  81.25801  &  1.58232  &  11.720 $\pm$ 0.030  &  11.865 $\pm$ 0.030  &  11.562 $\pm$ 0.032  &  11.825 $\pm$ 0.036  &  \nodata $\pm$ \nodata  &      2       &  CIII \\
931  &  05250205+0137210  &  81.25855  &  1.62252  &  11.435 $\pm$ 0.030  &  11.228 $\pm$ 0.030  &  11.326 $\pm$ 0.032  &  11.137 $\pm$ 0.033  &  \nodata $\pm$ \nodata  &      1       &  disk[8]? \\
948  &  05250362+0144121  &  81.26511  &  1.73670  &  11.949 $\pm$ 0.030  &  12.061 $\pm$ 0.030  &  11.831 $\pm$ 0.034  &  11.876 $\pm$ 0.036  &  \nodata $\pm$ \nodata  &      2       &  CIII \\
\enddata  
\tablecomments{Table \ref{t:mem1a} is published in its entirety in the electronic edition of the {\it Astrophysical Journal}. 
A portion is shown here for guidance regarding its form and content.}
\tablenotetext{~}{References in column 10: 1 \citet{briceno05}; 2 \citet{briceno07b}; 3 \citet{briceno07c}}
\tablenotetext{~}{Disk types: CIII:disk less stars; CII:optically thick disks; EV:evolved disks; TD: transitional disk candidates; disk[8]?:excess at 8 \micron but no MIPS detections.}
\end{deluxetable}.

\begin{deluxetable}{llllllllllllll}
\rotate
\tabletypesize{\tiny}
\tablewidth{0pt}

\tablecaption{Members of the OB1b field \label{t:mem1b}}
\tablehead{
\colhead{OB1b} & \colhead {2MASS} & \colhead{RA(2000)} & \colhead{DEC(2000)} & \colhead{[3.6]} & \colhead{[4.5]} & \colhead{[5.8]} & \colhead{[8.0]} & \colhead{[24.0]} & \colhead{Ref} & \colhead{Disk}\\
\colhead{ID} & \colhead{ID} & \colhead{deg} & \colhead{deg} & \colhead{mag} & \colhead{mag} & \colhead{mag} & \colhead{mag} & \colhead{mag} & \colhead{opt} & \colhead{type} \\
}
\startdata
21  &  05290540-0127500  &  82.27250  &  -1.46390  &  11.830 $\pm$ 0.030  &  11.865 $\pm$ 0.030  &  11.787 $\pm$ 0.034  &  11.846 $\pm$ 0.038  &  \nodata $\pm$ \nodata  &	3	&  CIII \\
31  &  05290635-0152122  &  82.27648  &  -1.87008  &  12.751 $\pm$ 0.030  &  12.707 $\pm$ 0.031  &  12.632 $\pm$ 0.040  &  12.595 $\pm$ 0.057  &  \nodata $\pm$ \nodata  &	3	&  CIII \\
63  &  05290882-0125393  &  82.28679  &  -1.42760  &  11.268 $\pm$ 0.030  &  11.291 $\pm$ 0.030  &  11.221 $\pm$ 0.032  &  11.076 $\pm$ 0.033  &  \nodata $\pm$ \nodata  &	1	&  CIII \\
70  &  05290925-0121227  &  82.28856  &  -1.35633  &  12.383 $\pm$ 0.030  &  12.291 $\pm$ 0.031  &  12.122 $\pm$ 0.035  &  11.813 $\pm$ 0.038  &  \nodata $\pm$ \nodata  &	3	&  CIII \\
78  &  05291078-0117281  &  82.29495  &  -1.29115  &  12.004 $\pm$ 0.030  &  11.989 $\pm$ 0.030  &  11.906 $\pm$ 0.034  &  11.974 $\pm$ 0.040  &  \nodata $\pm$ \nodata  &	3	&  CIII \\
89  &  05291202-0112236  &  82.30010  &  -1.20657  &  13.138 $\pm$ 0.031  &  13.085 $\pm$ 0.031  &  13.120 $\pm$ 0.052  &  13.067 $\pm$ 0.089  &  \nodata $\pm$ \nodata  &	3	&  CIII \\
148  &  05291821-0204066  &  82.32590  &  -2.06852  &  11.267 $\pm$ 0.030  &  11.152 $\pm$ 0.030  &  11.058 $\pm$ 0.032  &  10.917 $\pm$ 0.034  &  \nodata $\pm$ \nodata  &	3	&  CIII \\
209  &  05292326-0125153  &  82.34693  &  -1.42092  &  9.457 $\pm$ 0.030  &  9.238 $\pm$ 0.030  &  8.809 $\pm$ 0.030  &  8.392 $\pm$ 0.030  &  5.23 $\pm$ 0.03  &       1       &  CII \\
\enddata
\tablecomments{Table \ref{t:mem1b} is published in its entirety in the electronic edition of the {\it Astrophysical Journal}. 
A portion is shown here for guidance regarding its form and content.}

\tablenotetext{~}{References in column 10: 1 \citet{briceno05}; 2 \citet{briceno07b}; 3 \citet{briceno07c}}
\tablenotetext{~}{Disk types: CIII:disk less stars; CII:optically thick disks; EV:evolved disks; TD: transitional disk candidates; disk[8]?:excess at 8 \micron but no MIPS detections.}
\end{deluxetable}

\begin{deluxetable}{llllllllllllll}
\rotate
\tabletypesize{\tiny}
\tablewidth{0pt}
\tablecaption{Photometric candidates of the 25 Orionis aggregate \label{t:phot1a}}
\tablehead{
\colhead{OB1a} & \colhead {2MASS} & \colhead{RA(2000)} & \colhead{DEC(2000)} & \colhead{[3.6]} & \colhead{[4.5]} & \colhead{[5.8]} & \colhead{[8.0]} & \colhead{[24.0]} & \colhead{V} & \colhead{VI} & \colhead{Disk}\\
\colhead{ID} & \colhead{ID} & \colhead{deg} & \colhead{deg} & \colhead{mag} & \colhead{mag} & \colhead{mag} & \colhead{mag} & \colhead{mag} & \colhead{mag} & \colhead{mag} & \colhead{type} \\
}
\startdata
146 & 05230905+0125355 & 80.78774 & 1.42654 & 11.383 $\pm$ 0.030 & 11.355 $\pm$ 0.030 & 11.278 $\pm$ 0.032 & 11.284 $\pm$ 0.034 & \nodata $\pm$ \nodata & 15.75 $\pm$ 0.04 & 2.43 $\pm$ 0.05 & CIII \\
297 & 05233109+0144079 & 80.87958 & 1.73555 & 11.005 $\pm$ 0.030 & 11.079 $\pm$ 0.030 & 11.028 $\pm$ 0.031 & 10.967 $\pm$ 0.032 & \nodata $\pm$ \nodata & 13.91 $\pm$ 0.03 & 1.06 $\pm$ 0.05 & CIII \\
359 & 05234182+0152261 & 80.92428 & 1.87394 & 13.144 $\pm$ 0.031 & 13.088 $\pm$ 0.031 & 12.986 $\pm$ 0.041 & 13.039 $\pm$ 0.055 & \nodata $\pm$ \nodata & 19.11 $\pm$ 0.05 & 3.09 $\pm$ 0.09 & CIII \\
427 & 05235215+0136314 & 80.96732 & 1.60873 & 11.103 $\pm$ 0.030 & 11.191 $\pm$ 0.030 & 11.155 $\pm$ 0.032 & 11.074 $\pm$ 0.032 & \nodata $\pm$ \nodata & 13.84 $\pm$ 0.03 & 0.98 $\pm$ 0.05 & CIII \\
458 & 05235854+0151255 & 80.99394 & 1.85709 & 12.114 $\pm$ 0.030 & 12.047 $\pm$ 0.030 & 12.003 $\pm$ 0.034 & 12.043 $\pm$ 0.040 & \nodata $\pm$ \nodata & 17.29 $\pm$ 0.04 & 2.58 $\pm$ 0.07 & CIII \\
477 & 05240118+0128236 & 81.00493 & 1.47324 & 12.316 $\pm$ 0.030 & 12.289 $\pm$ 0.031 & 12.123 $\pm$ 0.035 & 12.165 $\pm$ 0.041 & \nodata $\pm$ \nodata & 16.87 $\pm$ 0.04 & 2.49 $\pm$ 0.06 & CIII \\
543 & 05241034+0155024 & 81.04312 & 1.91735 & 10.322 $\pm$ 0.030 & 10.428 $\pm$ 0.030 & 10.365 $\pm$ 0.031 & 10.309 $\pm$ 0.031 & \nodata $\pm$ \nodata & 13.16 $\pm$ 0.03 & 0.80 $\pm$ 0.05 & CIII \\
1626 & 05265473+0144337 & 81.72806 & 1.74271 & 10.168 $\pm$ 0.030 & 10.273 $\pm$ 0.030 & 10.061 $\pm$ 0.031 & 10.069 $\pm$ 0.031 & 9.83 $\pm$ 0.06 & 13.43 $\pm$ 0.03 & 0.86 $\pm$ 0.05 & EV \\
\enddata
\tablecomments{Table \ref{t:phot1a} is published in its entirety in the electronic edition of the {\it Astrophysical Journal}. 
A portion is shown here for guidance regarding its form and content.}
\tablenotetext{~}{Disk types: CIII:disk less stars; CII:optically thick disks; EV:evolved disks; TD: transitional disk candidates; disk[8]?:excess at 8 \micron but no MIPS detections.}
\end{deluxetable}

\begin{deluxetable}{llllllllllllll}
\rotate
\tabletypesize{\tiny}
\tablewidth{0pt}
\tablecaption{Photometric candidates of the OB1b field \label{t:phot1b}}
\tablehead{
\colhead{OB1b} & \colhead {2MASS} & \colhead{RA(2000)} & \colhead{DEC(2000)} & \colhead{[3.6]} & \colhead{[4.5]} & \colhead{[5.8]} & \colhead{[8.0]} & \colhead{[24.0]} & \colhead{V} & \colhead{VI} & \colhead{Disk}\\
\colhead{ID} & \colhead{ID} & \colhead{deg} & \colhead{deg} & \colhead{mag} & \colhead{mag} & \colhead{mag} & \colhead{mag} & \colhead{mag} & \colhead{mag} & \colhead{mag} & \colhead{type} \\
}
\startdata
74 & 05290984-0208250 & 82.29101 & -2.14030 & 12.269 $\pm$ 0.030 & 12.217 $\pm$ 0.031 & 12.141 $\pm$ 0.036 & 12.113 $\pm$ 0.042 & \nodata $\pm$ \nodata & 18.26 $\pm$ 0.05 & 2.89 $\pm$ 0.09 & CIII \\
187 & 05292160-0201546 & 82.34002 & -2.03184 & 11.142 $\pm$ 0.030 & 11.121 $\pm$ 0.030 & 11.073 $\pm$ 0.031 & 11.025 $\pm$ 0.034 & \nodata $\pm$ \nodata & 15.91 $\pm$ 0.06 & 2.04 $\pm$ 0.09 & CIII \\
203 & 05292300-0126559 & 82.34584 & -1.44888 & 13.500 $\pm$ 0.131 & 13.262 $\pm$ 0.032 & 13.186 $\pm$ 0.042 & 13.184 $\pm$ 0.085 & \nodata $\pm$ \nodata & 18.89 $\pm$ 0.05 & 2.83 $\pm$ 0.08 & CIII \\
207 & 05292313-0149203 & 82.34638 & -1.82231 & 13.109 $\pm$ 0.031 & 13.105 $\pm$ 0.031 & 13.059 $\pm$ 0.051 & 13.451 $\pm$ 0.153 & \nodata $\pm$ \nodata & 18.83 $\pm$ 0.05 & 2.63 $\pm$ 0.08 & CIII \\
222 & 05292426-0207354 & 82.35112 & -2.12651 & 12.953 $\pm$ 0.031 & 13.055 $\pm$ 0.031 & 12.730 $\pm$ 0.042 & 12.825 $\pm$ 0.063 & \nodata $\pm$ \nodata & 17.78 $\pm$ 0.05 & 2.27 $\pm$ 0.08 & CIII \\
236 & 05292591-0144580 & 82.35797 & -1.74945 & 10.810 $\pm$ 0.030 & 10.832 $\pm$ 0.030 & 10.757 $\pm$ 0.031 & 10.707 $\pm$ 0.032 & \nodata $\pm$ \nodata & 14.64 $\pm$ 0.04 & 1.41 $\pm$ 0.06 & CIII \\
276 & 05293010-0114446 & 82.37544 & -1.24573 & 11.116 $\pm$ 0.030 & 11.217 $\pm$ 0.030 & 11.175 $\pm$ 0.032 & 11.107 $\pm$ 0.034 & \nodata $\pm$ \nodata & 14.37 $\pm$ 0.00 & 1.25 $\pm$ 0.00 & CIII \\
283 & 05293049-0121500 & 82.37707 & -1.36389 & 11.303 $\pm$ 0.030 & 11.376 $\pm$ 0.030 & 11.314 $\pm$ 0.032 & 11.309 $\pm$ 0.036 & \nodata $\pm$ \nodata & 14.56 $\pm$ 0.04 & 1.30 $\pm$ 0.06 & CIII \\
342 & 05293578-0148046 & 82.39909 & -1.80129 & 11.138 $\pm$ 0.030 & 11.185 $\pm$ 0.030 & 11.129 $\pm$ 0.031 & 11.144 $\pm$ 0.034 & \nodata\enddata
\tablecomments{Table \ref{t:phot1b} is published in its entirety in the electronic edition of the {\it Astrophysical Journal}. 
A portion is shown here for guidance regarding its form and content.}
\tablenotetext{~}{Disk types: CIII:disk less stars; CII:optically thick disks; EV:evolved disks; TD: transitional disk candidates; disk[8]?:excess at 8 \micron but no MIPS detections.}
\end{deluxetable}

\clearpage

\end{document}